\documentclass{epl}
\usepackage[english]{babel}
\usepackage{amssymb}	

\title{Electrostatic colloid-membrane complexation}
\author{Christian C. Fleck, Roland R. Netz}
\institute{
  \inst{1} Fachbereich Physik, Universit\"at Konstanz, Universit\"atsstrasse 10, 78457 Konstanz, Germany\\
  \inst{2} Sektion Physik, LMU Munich, Theresienstrasse 37, 80333 Munich, Germany
}

\begin{document}
\bibliographystyle{/usr/local/teTeX/share/texmf.tetex/bibtex/bst/misc/phjcp}
\maketitle
\begin{abstract}
We investigate numerically and on the scaling level the adsorption of a charged colloid on an oppositely charged flexible membrane. We show that the long ranged character of the electrostatic interaction leads to a wrapping reentrance of the complex as the salt concentration is varied. The membrane wrapping depends on the size of the colloid and on the salt concentration and only for intermediate salt concentration and colloid sizes we find full wrapping. From the scaling model we derive simple relations for the phase boundaries between the different states of the complex, which agree well with the numerical minimization of the free energy.  
\end{abstract}

When a charged spherical particle adsorbs on an oppositely charged flexible membrane, the membrane deforms and wraps 
around the particle. This mechanism plays an important role in many physical and biological processes. 
Introduction of genes into cells via cationic lipid-DNA complexes has been one important  theme in biological research over the last few decades. Recent investigations show that endocytosis of lipid-DNA-complexes is triggered by electrostatic interactions with the cell membrane \cite{felgner-jbc94,ahearn-gtmb99,lin-bpj03}. Moreover, some viruses rely on internalization via endocytic uptake, where non-specific electrostatic interactions (and not a particular structural motif) seem to play a crucial role \cite{stegmann-jbc86,poranen-jcb99,kwong-jv00,cooper-bba03}.  In all cases the transfection efficiency depends on the size of the interacting particle \cite{xu-bpj99,smisterova-jbc01}, on the charge densities of the cell and particle surfaces and on the salt concentration \cite{ahearn-gtmb99,lin-bpj03,stegmann-jbc86,xu-bpj99,hui-bpj96}. 
But also in many physical experiments wrapping of  particles by membranes is important \cite{dietrich-jpf97,dimova-epjb99, koltover-prl99}.
Motivated by these examples, we investigate in this paper the role of the electrostatic interaction between a charged colloidal particle and an oppositely charged membrane in the adsorption and wrapping process.  The adsorption of a neutral spherical particle on a neutral fluid membrane due to a local adhesion potential has been studied theoretically already in \cite{deserno-epl03}, discussing the role of the elastic parameters of the membrane in the wrapping process. Furthermore, the wrapping of a colloid by a membrane is related to the budding of vesicles, which was subject to many studies \cite{juelicher-prl93,seifert-aip97}.
In the electrostatic case considered in this paper, the attraction between the colloid and the membrane as well as the repulsion between the membrane segments are shown to play a crucial role. We find that it is the long-ranged character of the electrostatic interaction, which  leads to a reentrance phenomena depending on the salt concentration: In case of low salt concentration or small colloids, the electrostatic interaction is barely screened and due to membrane-membrane repulsion the membrane only slightly wraps around the colloid. By increasing the salt concentration or the size of the colloid the screening of the membrane-membrane repulsion leads to a wrapping of the colloid by the membrane. Increasing the salt concentration or the size further, the electrostatic attraction between the colloid and the membrane becomes too weak compared with the mechanical bending energy of the membrane and again the membrane is only slightly wrapped. Therefore, the membrane wraps the colloid optimally only for intermediate salt concentrations or colloid sizes. This is similar to the complexation  behavior of a polyelectrolyte and an oppositely charged sphere as discussed in \cite{netz-macromole99}.

In our model, the membrane is described as a two dimensional elastic surface with homogeneous charge density $\sigma$, bending rigidity $K_0$ and surface tension $\gamma$; the colloid is approximated by a homogeneously charged sphere with radius $R$ and charge $Z$. We describe the electrostatic interactions by Debye-H\"uckel (DH) potentials. The simplest free energy to describe the system considered here, in units of $k_B T$, is given by the sum of the following three parts \cite{statmechmembranes}:
\begin{eqnarray}
\label{eq_hamiltonian_mech}
F_{\mathit{mech}}&=&\frac{K_0}{2}\int d^2\mathbf{\rho}\sqrt{g(\mathbf{\rho})}\left(\Delta \mathbf{X}(\mathbf{\rho}\right))^2 +\gamma_{\mathit{eff}}\int d^2\mathbf{\rho}\sqrt{g(\mathbf{\rho})}\\
\label{eq_hamiltonian_rep}
F_{\mathit{rep}}&=&\frac{\ell_B\sigma^2}{2}\int d^2\mathbf{\rho}\int d^2\mathbf{\rho}\prime\sqrt{g(\mathbf{\rho})g(\mathbf{\rho}\prime)}\,\frac{\exp\left\{-\kappa |\mathbf{X}(\mathbf{\rho})-\mathbf{X}(\mathbf{\rho}\prime)|\right\}}{| \mathbf{X}(\mathbf{\rho})-\mathbf{X}(\mathbf{\rho}\prime))|}\\
\label{eq_hamiltonian_attr}
F_{\mathit{attr}}&=&-\frac{\ell_B Z \sigma e^{\kappa R}}{1+\kappa R}\int d^2\mathbf{\rho}\sqrt{g(\mathbf{\rho})}\,\frac{\exp\left\{-\kappa |\mathbf{X}(\mathbf{\rho})|\right\}}{| \mathbf{X}(\mathbf{\rho})|}
\end{eqnarray}

The vector $\mathbf{X}(\mathbf{\rho})$ describes the position of the membrane in space, where $\rho=(\rho_1,\rho_2)$ are the internal coordinates of the membrane. $g(\mathbf{\rho})$ is the determinant of the induced metric field defined by $g_{ij}=\partial_i\mathbf{X}\partial_j\mathbf{X}$ and $\Delta=g^{-\frac{1}{2}}\partial_i\left(\sqrt{g}g^{ij}\partial_j\right)$ is the Laplace-Beltrami operator on the surface. The purely mechanical (elastic) contribution to the free energy $F_{\mathit{mech}}$ consists of the bending energy (first term) and the surface tension (second term). The repulsive electrostatic interaction between the membrane segments is given by $F_{\mathit{rep}}$ and the electrostatic membrane-colloid attraction is denoted by $F_{\mathit{attr}}$. The inverse Debye-H\"uckel screening length is given by $\kappa:=\sqrt{8\pi\ell_B c}$ ($c$ is the salt concentration), while the Bjerrum length $\ell_B:=e^2/4\pi \varepsilon_0\varepsilon k_B T$ measures the distance at which two elementary charges interact with thermal energy  $k_B T$. 
The membrane is fixed at a circular frame of radius $L$ connected to an area reservoir. In this way the projected area of the membrane remains constant as the membrane wraps around the colloid. In particular, this means that also the total charge of the system is not preserved, which generates an electrostatic tension $\gamma_{\mathit{el}}=-\pi\ell_B\sigma^2/\kappa$. Therefore the effective tension $\gamma_{\mathit{eff}}=\gamma+\gamma_{\mathit{el}}$ appearing in eq.(\ref{eq_hamiltonian_mech}) consists of the sum of the bare tension $\gamma$ and the electrostatic tension $\gamma_{\mathit{el}}$. 

\begin{figure}[top]
\begin{minipage}{72 mm}
{\bf a)}\quad$\tilde{Z}$\\
\onefigure[width=65 mm]{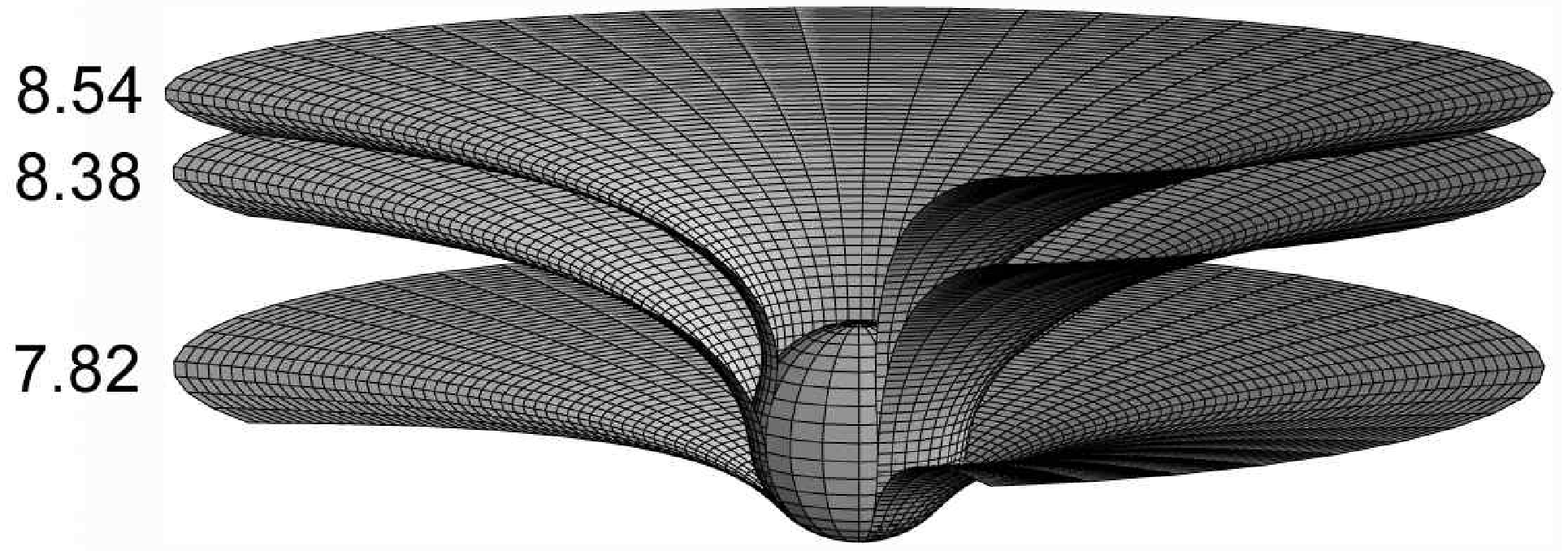}\\
{\bf b)}$\quad\tilde{\gamma}$\\
\onefigure[width=65 mm]{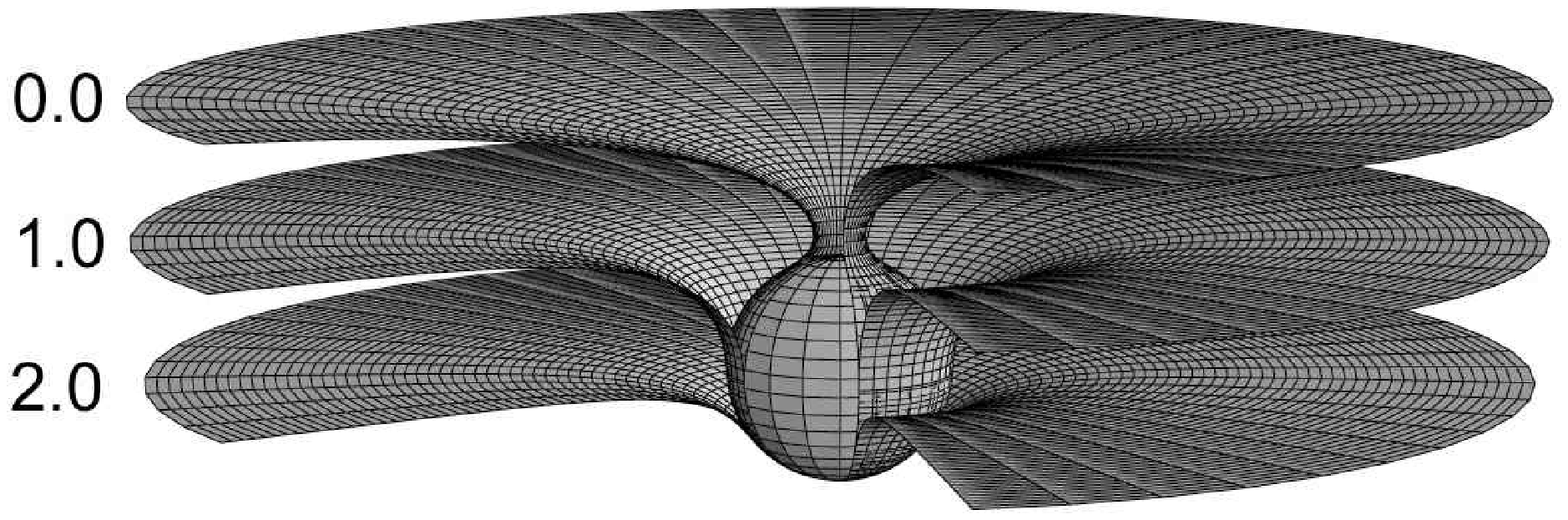}\\
{\bf c)}\\
\onefigure[width=65 mm]{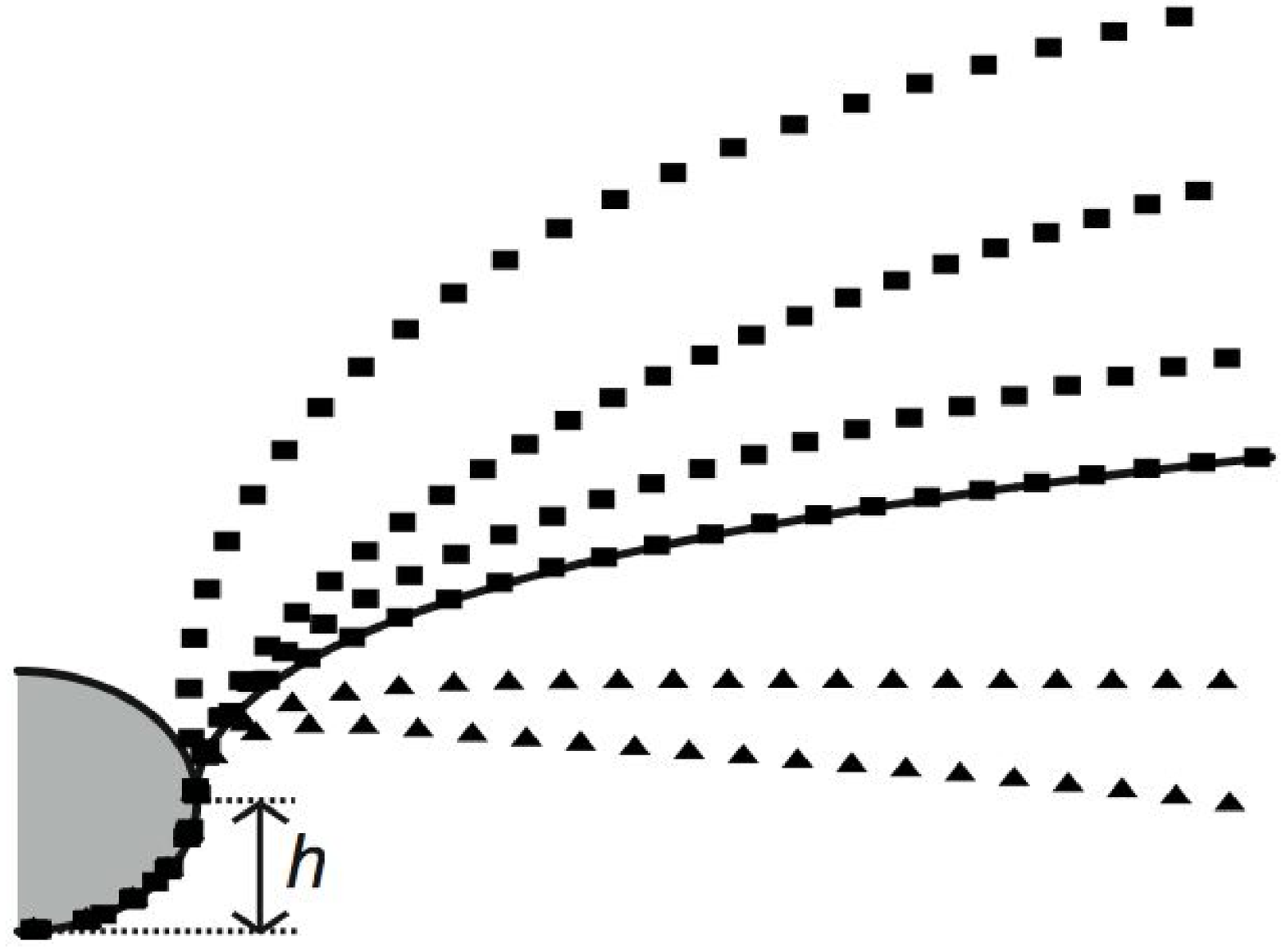}
\end{minipage}
\hfill
\begin{minipage}{72 mm}
{\bf d)}\quad$\tilde{Z}$\\
\onefigure[width=65 mm]{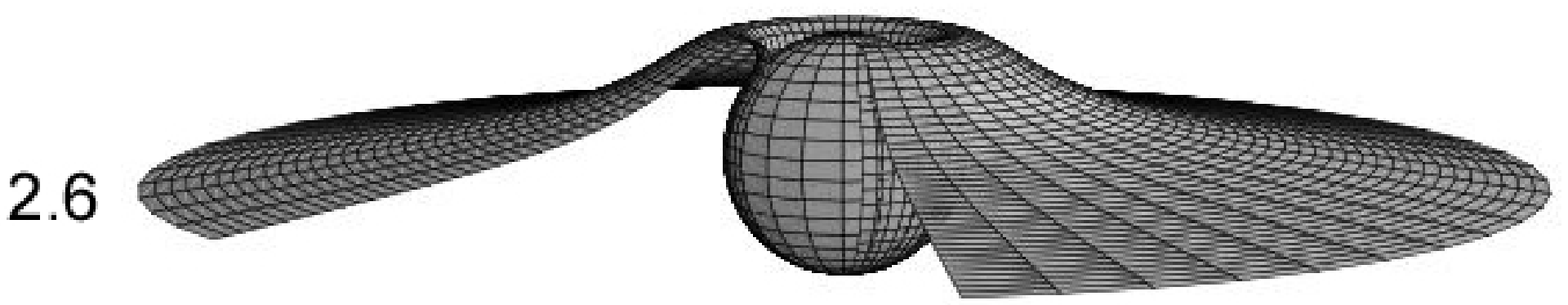}
\vspace{0.9cm}

\onefigure[width=65 mm]{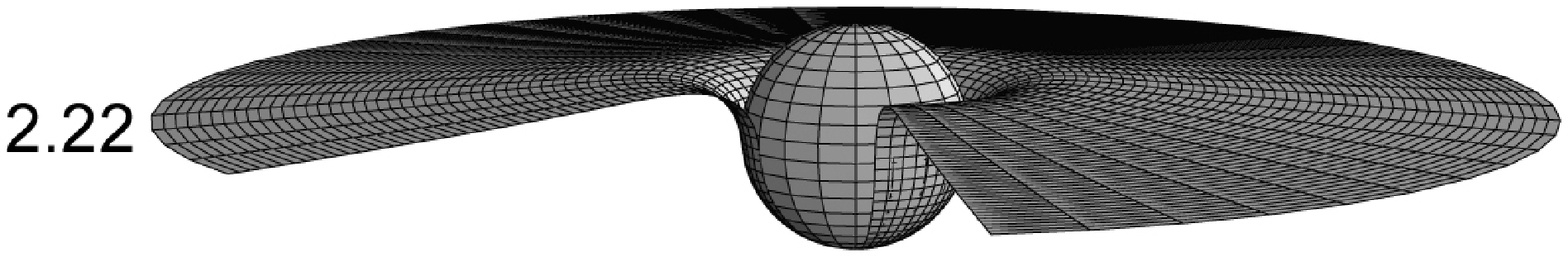}
\vspace{0.9cm}

\onefigure[width=65 mm]{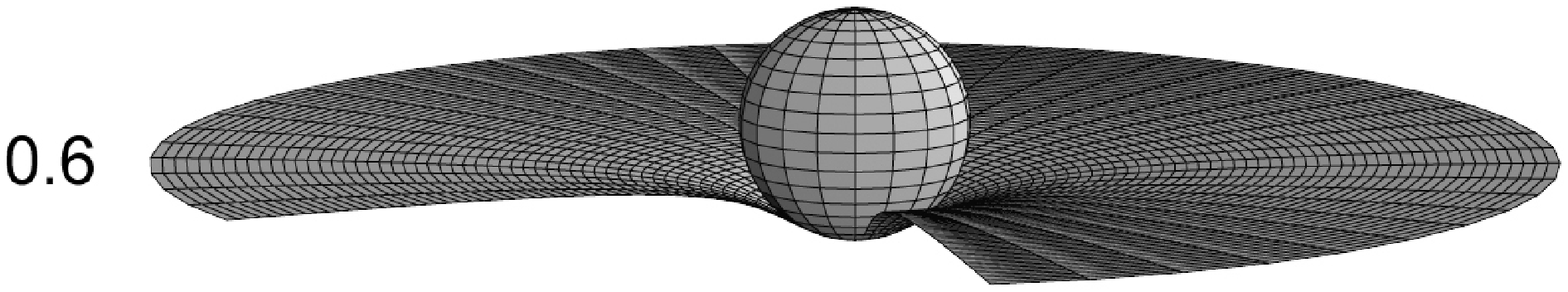}
\vspace{1.5cm}

{\bf e)}\\
\onefigure[width=65 mm]{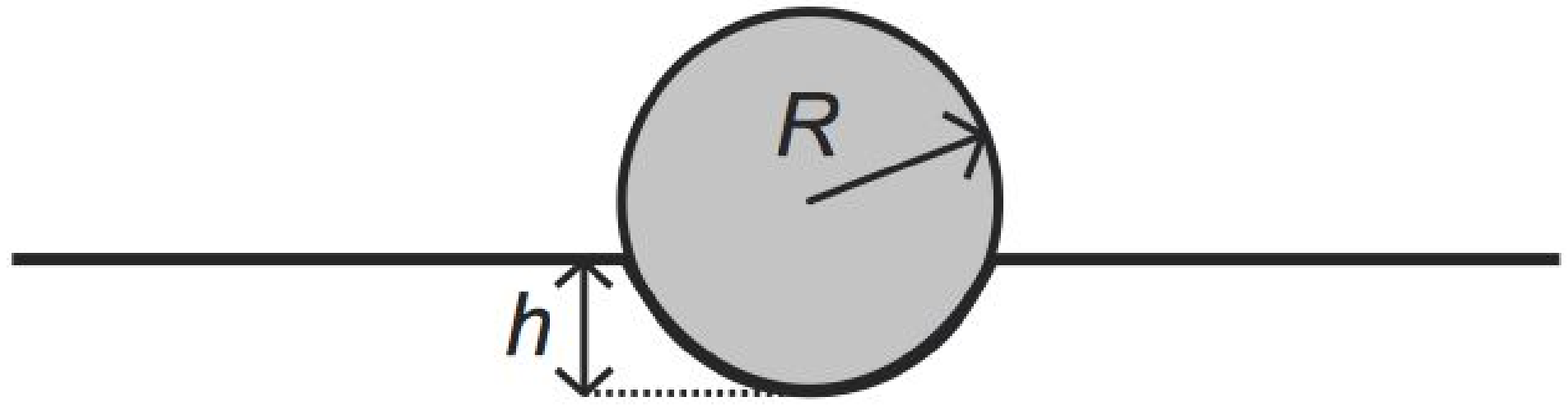}
\end{minipage}
\caption{\label{fig_shapes} Figure a, b, c and d show wrapping shapes of the membrane. 
{\bf a)}: High salt/large colloid limit ($\kappa R=50$) for $\tilde{\gamma}=0$ and $\tilde{\sigma}=1.0$. 
{\bf b)}: Unwrapping of the membrane by increasing the surface tension is shown ($\tilde{\sigma}=1.0$, $\kappa R=10.0$ and $\tilde{Z}=2.90$).
{\bf c)}: Increasing deviation from the cateniodal shape of the membrane shapes by decreasing $\kappa R$. The numerically obtained membrane surfaces (symbols) as well as the catenoid (solid line) detach from the sphere all at the same wrapping height ($h=R$). The parameters are: $\kappa R=45\,,20\,,10\,,2$ and $\tilde{Z}=7.46\,,2.8\,,1.1\,,0.52$ (squares, from bottom to top), $\kappa R=0.5,\,0.1$ and $\tilde{Z}=1.34,\,2.22$ (triangles, from top to bottom), $\tilde{\sigma}=1$ and $\tilde{\gamma}=0$ in all cases.
{\bf d)}: Low salt/small colloid limit ($\kappa R=0.1$) for $\tilde{\gamma}=0$ and $\tilde{\sigma}=1.0$.
{\bf e)}: For the scaling model we subdivide the membrane into a spherical segment, which touches the sphere and a planar ring.}
\end{figure}

We analyze the free energy of the system, $F=F_{\mathit{mech}}+F_{\mathit{rep}}+F_{\mathit{attr}}$, using two approaches. First, we minimize the free energy numerically and, second, we investigate the free energy in a simplified geometry on a scaling level.
We distinguish three different states of the complex, which can be defined as the membrane wraps around the colloid using the wrapping height $h$ defined by the point at which the membrane detaches from the colloid (see fig.\ref{fig_shapes}(c)):  i) the point contact state: the membrane touches the sphere only in one point ($h=0$), ii) the touching state: membrane touches the colloid over a finite area, but less then half of the colloid ($R>h>0$) iii) the wrapped state: the membrane covers more then half of the colloid surface ($2R\ge h\ge R$). Defining the wrapping  threshold at $h=R$ and not at $h=2R$ (as was done e.g. in \cite{deserno-epl03}) is necessary in order to avoid the singularity at $h=2R$ caused by the touching of charged surfaces in the numerical minimization.
In what follows, we rescale the free energy $\tilde{F}:=F/k_B T2\pi K_0$ and use rescaled variables: $\tilde{Z}:=Z/4\pi\sigma R^2$, $\tilde{\sigma}:=\sigma\sqrt{\ell_B R^3/K_0}$, $\tilde{\gamma}:=\gamma R^2/K_0$ and $\kappa R$.

\begin{figure}[top]
\begin{minipage}{72 mm}
{\bf a)}\\
\onefigure[width=65 mm,height=40mm]{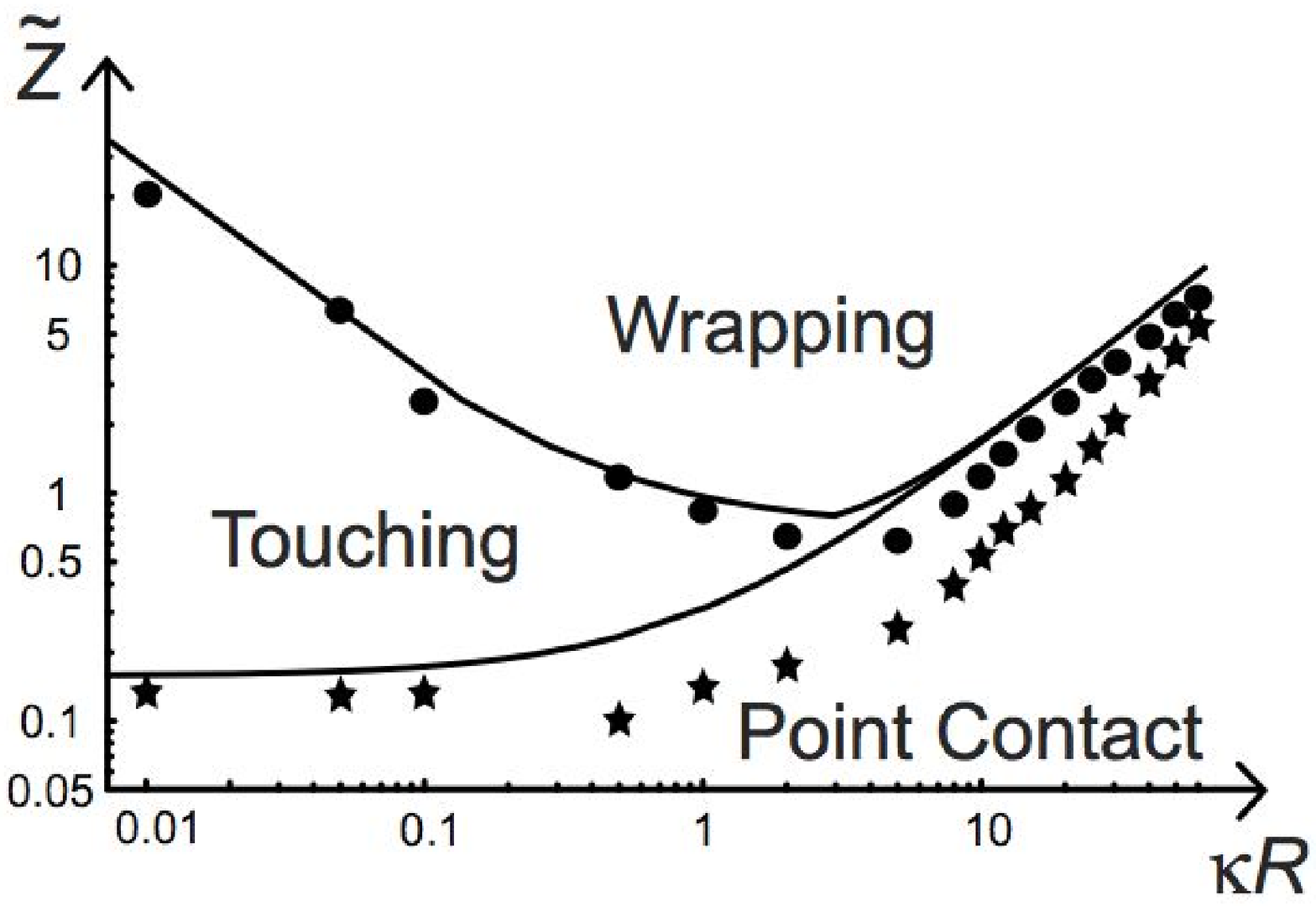}
\end{minipage}
\begin{minipage}{72 mm}
{\bf b)}\\
\onefigure[width=65 mm,height=40mm]{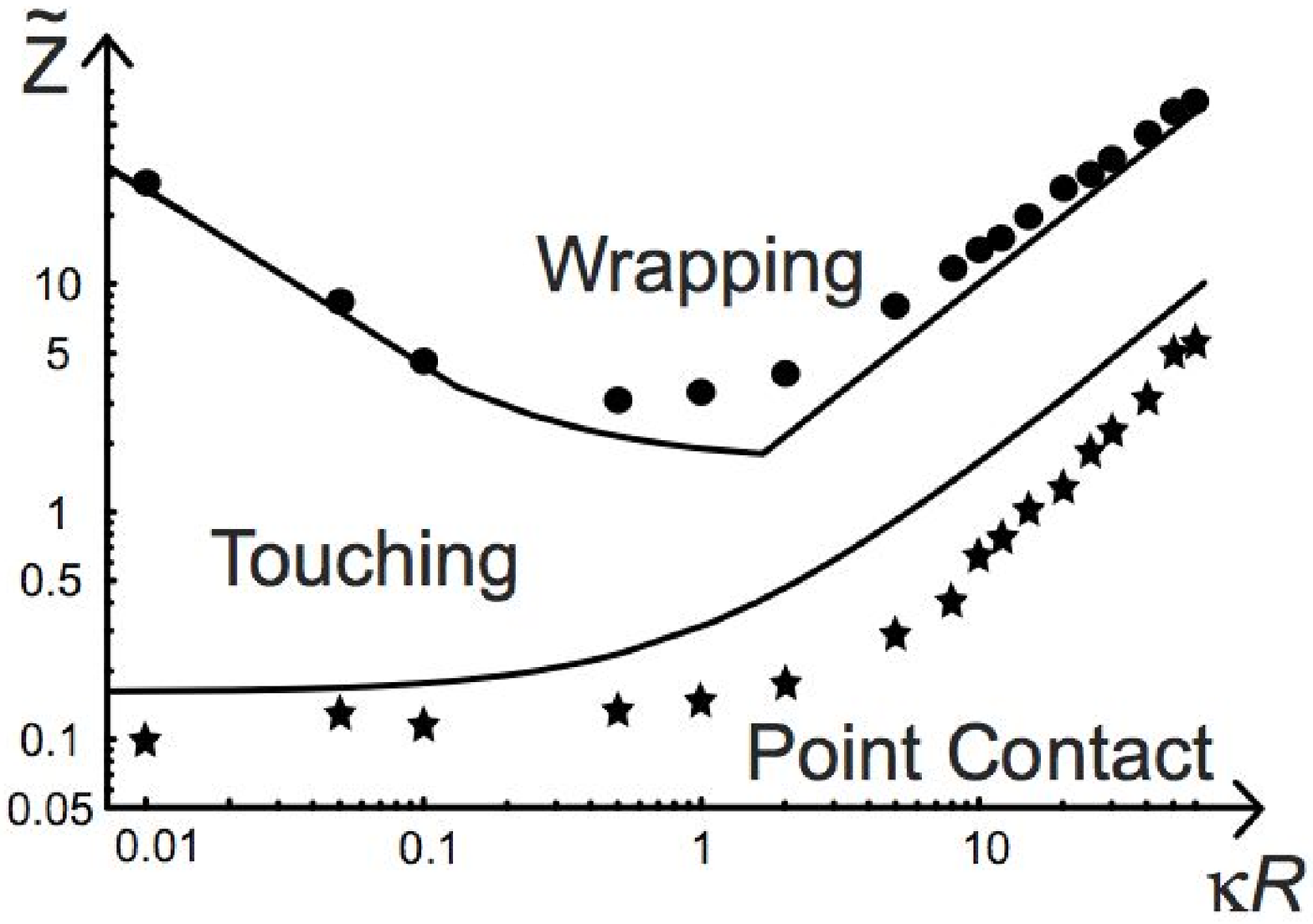}
\end{minipage}
\caption{\label{fig_phase_hom} Complexation phase diagrams for the membrane-colloid system as a function of $\tilde{Z}$ and $\kappa R$ on a logarithmic scale. In fig.({\bf a}) the surface tension is zero. The lower line denotes the touching transition, eq.(\ref{eq_touching_low_salt}), and the upper line the wrapping transition, eq.(\ref{eq_wrapping_low_salt}) and eq.(\ref{eq_wrapping_high_salt}) . The difference between the touching and the wrapping threshold is $\sim(\kappa R)^{-1}$ according to our scaling theory. Thus, the touching phase vanishes for large $\kappa R$. Fig.({\bf b}) shows the touching and the wrapping transition for finite surface tension, $\tilde{\gamma}=10$. In both figures the stars denote the results for touching threshold and the circles the results for wrapping threshold obtained from the full model.  The rescaled charge density on the membrane is in all cases given by $\tilde{\sigma}=1.0$.}
\end{figure}

In the numerical minimization of the free energy we use a frame with radius $L=2.5\pi R$ and model the hard-core repulsion between membrane and colloid by a shifted and truncated Lennard-Jones Potential, $v_{LJ}(\mathbf{r})=g[(R/|\mathbf{r}|)^{12}-2(R/|\mathbf{r}|)^{6}+1]$ for $|\mathbf{r}|\le R$ and $v_{LJ}(\mathbf{R})=0$ otherwise. We choose the parameter $g$ in such a way that the equilibrium membrane-sphere separation is $0.999 R$. Beside the constraint of constant projected area we impose, if not stated otherwise, the boundary condition that the membrane becomes planar at the edge. Fig.\ref{fig_shapes}(a) shows for three different colloid charges $\tilde{Z}$ the numerically obtained membrane shapes for $\kappa R=50$ (high salt/large colloid limit) and vanishing surface tension $\tilde{\gamma}$. For $\tilde{Z}=7.82$ the membrane is in the touching state. By elevating $\tilde{Z}$ the membrane bends further around the colloid and is for $\tilde{Z}=8.38$ right at the transition point between the touching and the wrapped state, while for $\tilde{Z}=8.54$ the membrane is in the wrapped state. Due to the strong screening of the electrostatic interaction the dominating energy contribution stems from the mechanical bending energy and the membrane wraps into a catenoid-like shape (surface of revolution of the curve $x(s)\approx \ln(s)$ for arc-length $s\gg 1$ \cite{osserman}), reducing in this way its elastic bending energy. In fig.\ref{fig_shapes}(b) we show the unwrapping of the membrane by increasing the surface tension for $\kappa R=10$ and $\tilde{\sigma}=1.0$ (the numbers denote $\tilde{\gamma}$). For zero surface tension the membrane is in the wrapped state. Increasing the surface tension the membrane unwraps and is for $\tilde{\gamma}=1$ right at the transition point to the touching state. By increasing the surface tension further to $\tilde{\gamma}=2.0$ the membrane goes into the touching state. In case of small $\kappa R$ (low salt concentration/small colloids), the electrostatic interactions are dominant and thus govern the shape of the membrane. In this case the long-ranged electrostatic attraction pulls the membrane closer to the colloid, producing the volcano-like shape presented in fig.\ref{fig_shapes}(d) for $\kappa R=0.1$ and $\tilde{\sigma}=1.0$ (the numbers denote $\tilde{Z}$). Comparison of the wrapped states in fig.\ref{fig_shapes}(a) and fig.\ref{fig_shapes}(d) demonstrates clearly the different wrapping structure of the membrane for small and large $\kappa R$, respectively. The shapes in fig.\ref{fig_shapes}(c) are obtained without imposing a boundary condition for the membrane at the edge in order to compare the wrapping shapes (symbols) of the membrane for different $\kappa R$ and $\tilde{Z}$ with a catenoid (solid line) touching the colloid at $h=R$. The rescaled colloid charge $\tilde{Z}$ was adjusted in such a way that the wrapping height of the membrane is in all cases the same ($h=R$). It can be clearly seen that the deviation from the catenoidal-shape becomes pronounced as $\kappa R$ decreases, since the range of the electrostatic interaction increases.
The calculation of membrane shapes for different parameters allows us to measure the complexation diagrams presented in fig.\ref{fig_phase_hom}. Before we discuss these diagrams we calculate the phase boundaries between the different states of the complex using a simplified scaling model.

In the geometry shown in fig.\ref{fig_shapes}(e) the calculation of the free energy is analytically tractable, yielding the free energy in form of a Landau function with the rescaled wrapping height $\tilde{h}:=h/R$ playing the role of the order parameter, from which we can extract simple scaling relations for the phase boundaries. The mechanical  energy and the electrostatic membrane-colloid attraction read:
\begin{eqnarray}
\label{eq_scaling_mech}
\tilde{F}_{\mathit{mech}}&=&2\tilde{h}+\frac{\tilde{\gamma}}{2}\tilde{h}^2\\
\label{eq_scaling_attr}
\tilde{F}_{\mathit{attr}}&=&-\frac{4\pi\tilde{\sigma}^2\tilde{Z}}{1+\kappa R}\tilde{h}\,.
\end{eqnarray}
For the electrostatic repulsion we find in the limit $\kappa R\ll 1$:
\begin{eqnarray}
\label{eq_scaling_rep_1}
\tilde{F}_{\mathit{rep}}&=&\frac{16\sqrt{2}\tilde{\sigma}^2}{3}\tilde{h}^{3/2} + \left(\frac{\pi\tilde{\sigma}^2}{2\kappa R}-\pi\tilde{\sigma}^2\right)\tilde{h}^2+\mathcal{O}(\kappa R,\tilde{h}^{5/2})
\end{eqnarray}
and in the limit $\kappa R\gg 1$:
\begin{eqnarray}
\label{eq_scaling_rep_2}
\tilde{F}_{\mathit{rep}}=\left\{\begin{array}{r@{\;:\;}l} 
\frac{16\sqrt{2}\tilde{\sigma}^2}{3}\tilde{h}^{3/2}+\left(\frac{\pi\tilde{\sigma}^2}{2\kappa R}-\pi\tilde{\sigma}^2-2\pi\kappa R\tilde{\sigma}^2\right)\tilde{h}^2+\mathcal{O}(\kappa R^{-2},\tilde{h}^{5/2}) &\sqrt{h}< \sqrt{2}/\kappa R\\
\frac{2\pi\tilde{\sigma}^2}{\kappa R^2}-\frac{\sqrt{2}\tilde{\sigma}^2}{\kappa R^2}h^{1/2}+\frac{\tilde{\sigma}^2}{4\sqrt{2}\kappa R^2}\tilde{h}^{3/2} +\frac{\tilde{\sigma}^2}{16\sqrt{2}\kappa R^2}\tilde{h}^{5/2}+\mathcal{O}(\kappa R^{-3},\tilde{h}^{3})& \sqrt{h} > \sqrt{2}/\kappa R
\end{array}\right .
\end{eqnarray}
Using the equations (\ref{eq_scaling_mech})-(\ref{eq_scaling_rep_2}) we can analyze the transitions between the different membrane states in the following way. The touching transition occurs when the coefficient in front of $\tilde{h}$ vanishes, yielding the exact result for the touching threshold:
\begin{eqnarray}
\label{eq_touching_low_salt}
\tilde{Z}_t=\frac{\kappa R+1}{2\pi\tilde{\sigma}^2} \,.
\end{eqnarray}
In the high salt limit the free energy given by the sum of the equations (\ref{eq_scaling_mech}), (\ref{eq_scaling_attr}) and (\ref{eq_scaling_rep_2}) exhibits a secondary minimum rendering the wrapping transition discontinuous. Note that this  discontinuity is of electrostatic origin and is different from the one found in \cite{deserno-epl03}. We determine the wrapping threshold $\tilde{Z}_w$ by resolving the system of equations $\tilde{F}(\tilde{h}_1)=\tilde{F}(\tilde{h}_2)$ and $\partial_{\tilde{h}}\tilde{F}(\tilde{h}_2)=0$ for $\tilde{Z}$ subject to the constraint $\tilde{h}_2\le 2$, where $\tilde{h}_1$ is the location of first and $\tilde{h}_2$ is the location of the second minimum. 
In the limit $\kappa R\ll 1$ the threshold reads:
\begin{eqnarray}
\label{eq_wrapping_low_salt}
\tilde{Z}_w=\frac{1}{4\kappa R}+\frac{2+\tilde{\gamma}}{4\pi\tilde{\sigma}^2}+\frac{2\sqrt{2}}{\pi}-\frac{1}{4}+\mathcal{O}(\kappa R),.
\end{eqnarray}
and in case of $\kappa R\gg 1$ we find:
\begin{eqnarray}
\label{eq_wrapping_high_salt}
\tilde{Z}_w=\frac{2+\tilde{\gamma}}{4\pi\tilde{\sigma}^2}\kappa R+\frac{2+\tilde{\gamma}}{4\pi\tilde{\sigma}^2}+\frac{8\pi-11}{32\kappa R}+\mathcal{O}\left((\kappa R)^{-2}\right)\,.
\end{eqnarray}

The resulting complexation diagram in the $\tilde{Z}/\kappa R$ plane is presented in fig.\ref{fig_phase_hom}, where the lines denote the scaling results eq. (\ref{eq_touching_low_salt})-(\ref{eq_wrapping_high_salt}) and the symbols refer to the numerical solution of the full model.
\begin{figure}[top]
\begin{minipage}{72 mm}
{\bf a)}\\
\onefigure[width=65 mm,height=40mm]{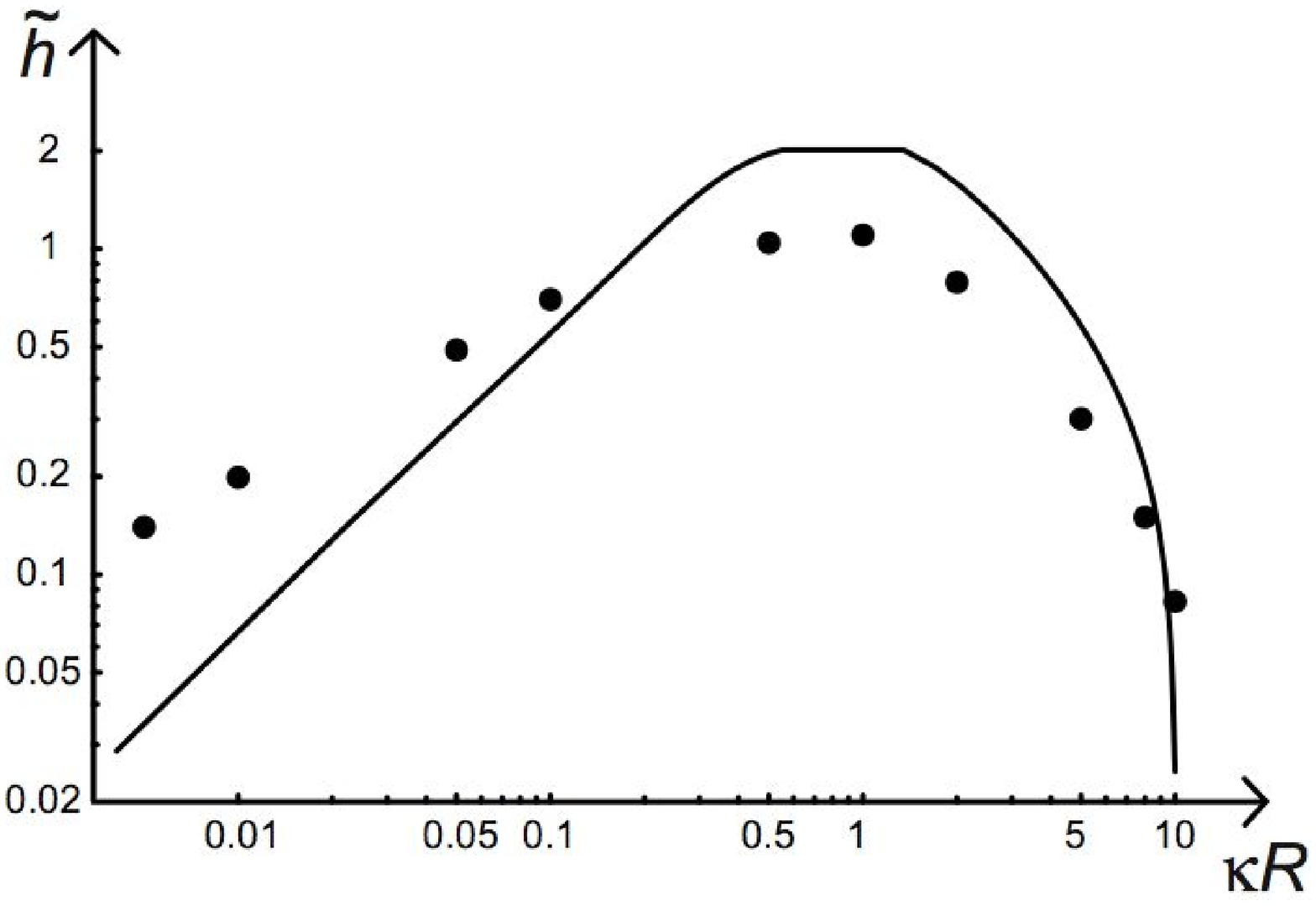}
\end{minipage}
\begin{minipage}{72 mm}
{\bf b)}\\
\onefigure[width=65 mm,height=40mm]{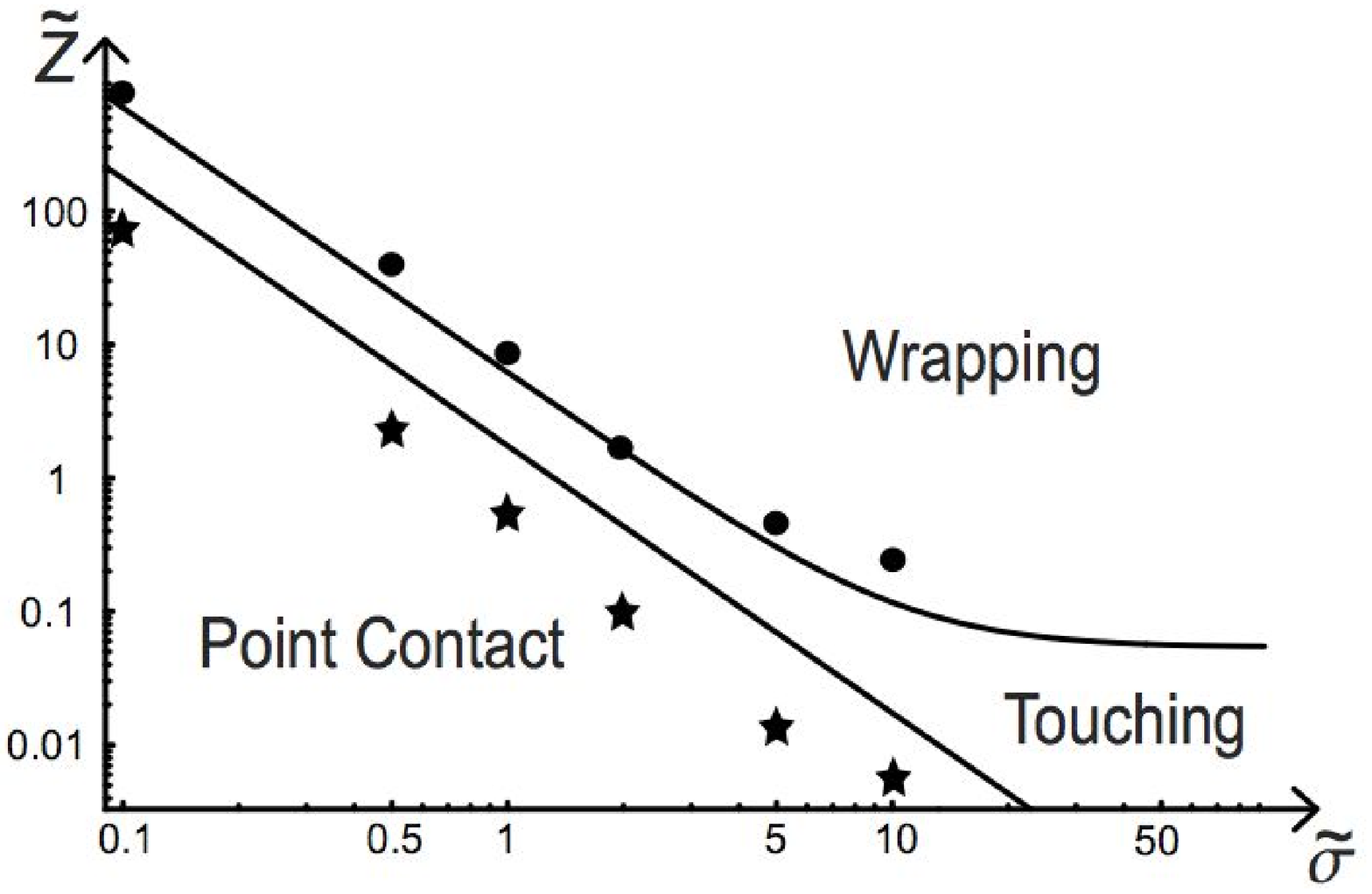}
\end{minipage}
\caption{\label{fig_phase_hom_sigma} {\bf a}): The wrapping height $\tilde{h}$ as a function of $\kappa R$. The line denote $\tilde{h}$ as obtained from the scaling model (eq.(\ref{eq_scaling_mech})-eq.(\ref{eq_scaling_rep_2})), while the solid circles denote the corresponding results obtained by numerical minimization of the full free energy. The parameters are: $\tilde{\sigma}=1.0$, $\tilde{Z}=2.2$ and $\tilde{\gamma}=5.0$. {\bf b}): the complexation phase diagram for the membrane-colloid system as a function of $\tilde{Z}$ and $\tilde{\sigma}$ is shown on a logarithmic scale ($\kappa R=10$, $\tilde{\gamma}=5$). The lower line denotes the touching transition and the upper line shows the wrapping transition. The stars symbolize the touching threshold and the solid circles the wrapping threshold as obtained from the numerical solution of the full model.}
\end{figure}
For $\kappa R$ small the touching transition depends in leading order only on the rescaled charge density $\tilde{\sigma}$, while in the high salt limit the touching threshold is $\tilde{Z}_t\sim\kappa R/\tilde{\sigma}^2$. If the surface tension is zero (fig.\ref{fig_phase_hom}(a)), the touching threshold differs from the wrapping threshold for $\kappa R\gg 1$ by a term of $\mathcal{O}\left((\kappa R)^{-1}\right)$. It follows from eq.(\ref{eq_touching_low_salt}) and eq.(\ref{eq_wrapping_high_salt}) that in the limit $\kappa R\to \infty$ the membrane goes for $Z\sigma\ell_B/\kappa=2 K_0$ directly from the point contact state with $\tilde{h}=0$ to the wrapping state with $\tilde{h}\ge 1$, skipping the touching region \cite{lipowsky-epl98, dinsmore-prl98}. This limit corresponds to the case of a local adhesion energy between a neutral colloid and neutral membrane and was discussed in \cite{deserno-epl03}. 
We find very good agreement between the phase boundaries derived from the scaling model and the values for touching and the wrapping threshold obtained from the full model, demonstrating that the simplified model captures the main physical properties of the system.

Starting from the touching state in fig.\ref{fig_phase_hom} and fixing $\tilde{Z}$ the complex goes by increasing the salt concentration into the wrapping state and by increasing the salt concentration further finally leaves the wrapping state and goes into the touching state again. This reentrance behavior of the complex can also be seen in fig.\ref{fig_phase_hom_sigma}(a), where we plot the wrapping height $\tilde{h}$, which minimizes the scaling free energy,  as a function of $\kappa R$. The solid line results from the scaling free energy, eq.(\ref{eq_scaling_mech})-(\ref{eq_scaling_rep_2}) with the constraint $\tilde{h}\le 2$ and the solid circles show the results obtained from the full model. The physical mechanism for this wrapping/unwrapping is the following. If $\kappa R$ is small, the electrostatic interaction is long ranged and the free membrane area is strongly self repelling. As the membrane bends around the colloid, charged membrane area is pulled closer together and  the resulting energy cost has to be paid for by increasing the membrane-colloid attraction. In particular, this means that for zero salt concentration ($\kappa R=0$), i.e. in the case of pure Coulomb interaction, the membrane only slightly bends around the colloid, no matter how large $\tilde{Z}$ is. On the other hand, if $\kappa R$ is large the electrostatic membrane-colloid attraction is screened and in order to overcome the bending energy of the membrane the rescaled colloid charge has to be increased again. 

When the bending rigidity $K_0$ is small and/or the surface charge density on the membrane is large the parameter $\tilde{\sigma}:=\sigma\sqrt{\ell_B R^3/K_0}$ becomes large. In this case the electrostatic interaction dominates the behavior of the complex. The limit $\tilde{\sigma}\to\infty$ corresponds to the case of vanishing bending rigidity and even an infinitesimal small charge on the sphere will result in a bending of the membrane, i.e. $\tilde{Z}_t\to 0$ as follows from eq.(\ref{eq_touching_low_salt}). On the opposite, the limit $\tilde{\sigma}\to0$ corresponds to infinitely large bending rigidity $K_0$ (stiff wall) or vanishing surface charge density $\sigma$ on the membrane and the membrane remains planar as long as $\tilde{Z}\tilde{\sigma}\to 0$, which also follows from eq.(\ref{eq_touching_low_salt}). In fig.\ref{fig_phase_hom_sigma}(b) we show the complexation diagram in the $\tilde{Z}/\tilde{\sigma}$-plane ($\kappa R=10$, $\tilde{\gamma}=5$). The symbols denote the results from the full model and the solid lines the results from the scaling model eq.(\ref{eq_touching_low_salt}) and eq.(\ref{eq_wrapping_high_salt}). Again, we find good agreement between the simple scaling model and the much more complex full model. The wrapping threshold diverges as $\tilde{Z}_w\sim\tilde{\sigma}^{-2}$ for $\tilde{\sigma}\to 0$, while for $\tilde{\sigma}\to\infty$ we find $\tilde{Z}_w\sim 1/\kappa R$, which follows from eq.(\ref{eq_wrapping_high_salt}). 

In this paper we have studied the role of the electrostatic interaction between a charged colloid and an oppositely charged fluid membrane in the wrapping process. For $\kappa R \ll 1$ the membrane wraps into a volcano-like shape, while in the opposite limit $\kappa R \gg 1$ the membrane wraps into a catenoid-like shape. We find a reentrance behavior in the wrapping process depending on $\kappa R$. For small as well as for large $\kappa R$ the membrane wraps the colloid only for large colloid charges. From the scaling model we derive simple relations for the phase boundaries of the complex, which agree well with our results from the more complex full  model.
The electrostatic repulsion between the membrane segments plays a crucial role in the adsorption process of charged colloid on an oppositely charged membrane. But also for the adhesion of neutral colloids at charged membranes in the case of a non-electrostatic local adhesion energy the membrane self repulsion is important. Since most bio-membranes are negatively charged the electrostatic mechanism studied in this paper plays certainly a role in many biological systems. \acknowledgments
Financial support by the "International Research Training Group Soft Condensed Matter" at the University of Konstanz, Germany, is acknowledged. 
\bibliography{/Users/chris/Documents/Physic/misc/litbank}
\end{document}